*Predicting the Stability of Geopolymer Activator Solutions for Optimised Synthesis through Thermodynamic Modelling*

Ramon Skane[a,b], Philip A. Schneider [c], William D.A. Rickard[b], Franca Jones [d], Arie van Riessen[b,d], Evan Jamieson[b,d], Xiao Sun[b]

[a] Corresponding Author, ramon.skane@postgrad.curtin.edu.au

[b] John de Laeter Centre, Curtin University, PO Box U 1987, Perth, WA 6102, Australia

[c] School of Engineering, The University of Western Australia, 35 Stirling Highway, 6009 Perth, Australia

[d] School of Molecular and Life Science, Curtin University, PO Box U 1987, Perth, WA 6102, Australia

[e] Future Battery Industries Cooperative Research Centre, The Hub Technology Park, Perth, WA 6102, Australia

### Highlights

1. Thermodynamic stability analysis supports the near-immediate preparation time of activator solutions ready for geopolymer production (batching) in a matter of minutes, which contrasts current practice whereby their synthesis is prepared over extensive, undefined and unqualified "equilibration" periods (sometimes spanning hours or days).

2. The dynamic thermochemical behaviour of geopolymer activator solutions (or "activators") has been quantified by an experimentally validated mathematical model that reflects laboratory experiments and could be extended to industrial applications.

3. The model and its results can be used to quantify the temperature, thermal stability and process control elements for developing quality-controlled activator solution systems for use in geopolymer synthesis.

4. An experimentally validated model quantifies system energy considerations, predicting the time course of the temperature of various activator solution reactions and their cooling periods, enabling better process control for the geopolymer practitioner.line n

**Abstract**

Geopolymers are an emerging class of binding materials used in sustainable cements, concretes, and composites. However, despite growing research, the lack of standardised processes and stability analyses for formulating activator solutions – a crucial component of geopolymer systems – remains a barrier to quality control and research advancement. This study presents an experimentally validated energy balance with thermodynamic phenomenon mathematically modelled for synthesising consistent geopolymer activator solutions. The model's general applicability enables dynamic assessments of user-specified systems, offering stability metrics for quality control in laboratory and industrial settings. Fundamentally, the mathematical model can be used towards batching optimisation under user-defined conditions where dissolution of geopolymer precursors can be maximised via solution preparation and batching optimisation. The model results quantify experimentally validated temperature dynamics, thermodynamic stability, and process design/batching optimisation, challenging traditional practices in the literature that rely on undefined equilibration periods. Key findings demonstrate that stable, ready-to-use activator solutions can be achieved in as little as 1 minute, compared to the typically used 24-hour batching periods. This research paves the way towards standardised activator solution preparation and supports the development of Standard Operating Procedures (SOPs) for geopolymer synthesis, promoting consistency and scalability in geopolymer technology.

# 1 Introduction

A geopolymer, otherwise known as an aluminosilicate inorganic polymer, is the amorphous polymeric material that is formed when a suitably reactive amorphous solid, most commonly an aluminosilicate powder, is exposed to a specially designed chemical primer [1]. In literature, this is usually referred to as the "activator solution". The aluminosilicate precursor material undergoes several physical and chemical changes during the formation reaction until a series of chemically bonded mineral "geo"-polymeric networks are formed [1, 2]. These geopolymer networks bind raw materials, offering economic and environmental benefits especially when utilising abundant, low-cost industrial by-products. They have been developed as cementitious binders to replace Ordinary Portland Cement (OPC), targeting the reduction of virgin material use and the high greenhouse gas (GHG) emissions of OPC, which accounts for 5–7% of global GHGs and is the highest per tonne manufactured product in the world. [3, 4]. Over the past few decades, geopolymer research has received significant attention with abundant research literature [5]. There have been several industrial cases where geopolymer-based cements and concretes have been found to achieve high compressive strengths (i.e. >50 MPa) [6, 7], allowed for industrial symbiosis opportunities with significant by-product encapsulation potential [8, 9], high thermal and fire resistance [10] and a reduced equivalent greenhouse gas emissions ranging from 47 – 64% [11, 12] when compared to OPC.

## 1.1 Geopolymer Activator Solutions

Despite the growing body of research on geopolymers, there remains a significant lack of standardisation in their manufacturing and synthesis processes. This absence of consistency raises concerns about the reproducibility and comparability of experimental results, particularly regarding the properties of resulting geopolymers. Many studies outline various approaches to mix design, feedstock analysis, and batching, yet these methods can lead to geopolymers with varying properties (e.g., compressive strength) even when identical feedstocks are used [13]. At a minimum, this inconsistency would also lead to the underperformance of geopolymers in terms of product quality assurance, quality control, and the preclusion of technical and economic product optimisation.

The challenges are compounded by the complex thermochemistry, reaction kinetics, and solubility of the "activator solution" — a critical component in geopolymer systems, alongside the aluminosilicate precursor.

However, there is no consensus on terminology, with the activator solution sometimes referred to as a "hardener," "alkaliniser," or, most commonly, an "activator solution" [14, 15]. These naming conventions arise from different fields and opposing theories of complex solution chemistry, its dynamics and its role in forming geopolymers [1, 16]. For simplicity, the term "activator solution" will be used throughout this study.

The activator solution can vary widely in composition and energetics depending largely on the raw material component feedstocks, implemented mix design and the mixing/batching procedure. All singular feedstocks in a geopolymer system, including those used to synthesise the activator solution, can be sourced as virgin or industrial by-product materials and also vary widely in composition, phase, solubility, crystallinity (if a solid-state), and the resultant chemical equilibria and speciation among other properties [17, 16]. In most applications presented in the literature, the activator solution is an alkaline mixture of soluble silicates or aluminates, an alkali metal hydroxide (e.g., NaOH or KOH), and water, with proportions determined by the mix design. However, the batching addition sequence of these feedstocks is also inconsistent throughout the literature [13]. In most cases, the activator solution is prepared by the sequenced addition of the alkali hydroxide to water, liberating energy due to the exothermic dissolution, followed by the addition of the sodium silicate or aluminate, depending on the system mix design. These feedstocks and their combinations are conceptualised generally in Figure 1, alongside an overview of the general geopolymer and activator solution system.

## 1.2 System Batching & Sequencing

In some cases, the alkali hydroxide solution is first prepared to a specific concentration (i.e. usually 6M, 8M, 10M, 12M or 14M NaOH) before adding other silicate/aluminate feedstocks, based on a general feedstock ratio (e.g. 'NaOH / $Na_2SiO_3$' weight ratio) [18, 19, 15, 20]. In other cases, the concentration is not fixed, varying depending on the molar analytes (e.g. molar Na or K, Si/Al, (Na or K)/(Si or Al) and $OH^-$) from the final activator solution to the geopolymer precursor making up the system [21, 22, 23]. In both cases, assuming there is no in-situ unstable precipitation event in the activator solution, the feedstock addition sequence is mostly unspecified in publications. Some researchers add the silicate/aluminate feedstock simultaneously with the alkali hydroxide dissolution, some after the exothermic solution has cooled to room temperature (which is often unspecified), and others only during geopolymer batching (i.e. mixing the activator solution with the aluminosilicate precursor). A practical example of the impact of this inconsistency is shown in Figure 2, where two activator solutions with identical compositions (i.e. 14.9 wt.% $SiO_2$, 20.3 wt.% $Na_2O$ and 64.8 wt.% $H_2O$) and ambient conditions were prepared using different addition sequences. Despite identical feedstocks, the resulting solutions differed significantly in their physiochemical properties. In Figure 2, the solution on the right precipitated into an unusable, gelatinous and non-stable slurry, with great difficulty in manual handling and solubilisation of geopolymer precursors, while the left solution remained relatively stable with lower viscosity and better handling properties. The only difference between these solutions was the addition sequence, highlighting how practitioners unaware of this variability may discard the unusable solution or use it with low confidence in quality and repeatability.

Further complicating matters is the lack of clarity surrounding the mixing period of the activator solution, which can range from minutes to hours or even "overnight" periods (which may mean a 24-hour or otherwise unspecified period) [24, 19, 25, 26]. These inconsistencies can lead to some mixtures unintentionally precipitating (or crystallising) immediately and others doing so unpredictably after an unknown period. In the case of the lengthier mixing periods (i.e. "overnight"), these procedures seem to serve as a subjectively empirical "equilibration" period without an assessment of the stability, viability and selection of an optimised activator solution for a given geopolymer system [22, 25, 27]. Whilst these literature methods have produced strong and durable geopolymers, the lack of quantitative analytics on the activator solution's stability to better assess its quality, consistency and impact in a geopolymer system is not well understood. At best, this misunderstanding yields issues for quality control purposes, inhibits the optimisation of geopolymer systems and underestimates the efficacy of activator solutions. At worst, this misunderstanding results in the discarding of 'precipitated' activator solutions that are otherwise usable – or even favourable – in synthesising geopolymers and precludes further research in the field.

This study focuses on thermochemical modelling of geopolymer activator solutions to quantitatively assess their stability. It also aims to provide insights for standardising synthesis procedures for activator solutions in geopolymer systems. The mathematical model developed is based on fundamental principles, making it applicable to a range of activator solutions with varying compositions, rather than relying on system-specific empirical equations [28]. This model supports the informed design of activator solutions and batching procedures, enabling the prediction of thermochemical dynamics to optimise geopolymer systems. Optimisation is partially achieved by maximising the theoretical reactivity of the geopolymer solid aluminosilicate precursor, which is linked to higher system temperatures that enhance the dissolution of precursor materials [25, 29]. Overall, this work seeks to standardise activator solution synthesis, providing a foundation for more reliable, reproducible and efficient geopolymer production.

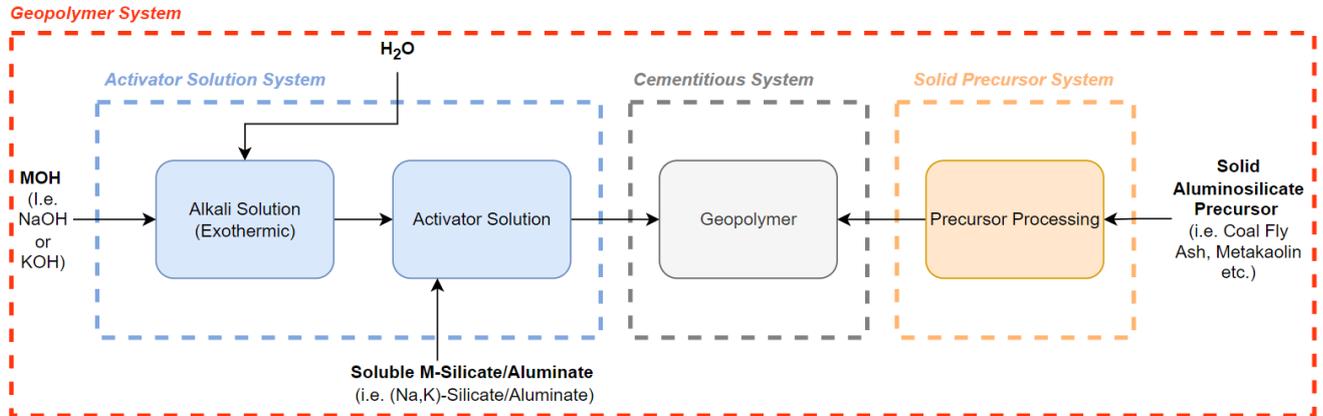

**Figure 1:** Process Flow Diagram illustrating common geopolymer systems comprising component activator solution and solid precursor systems.

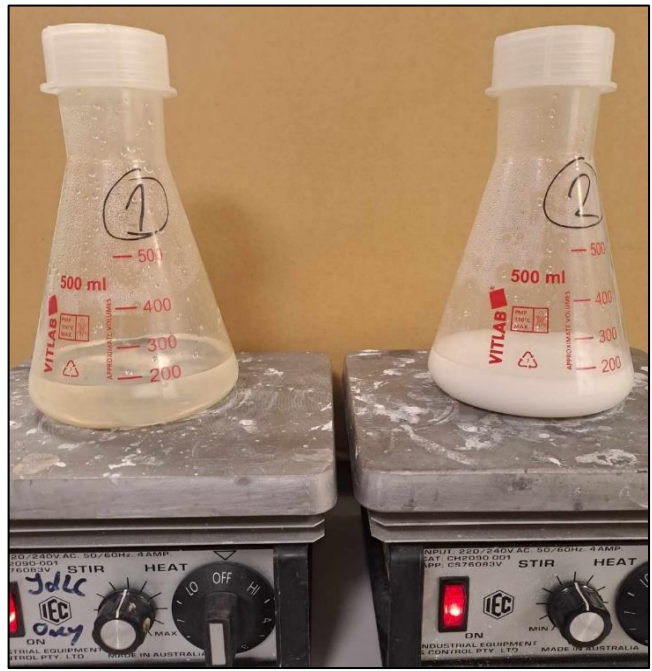

**Figure 2:** Two activator solutions with identical initial feedstock compositions of 14.9 wt.% $SiO_2$, 20.3 wt.% $Na_2O$ and 64.8 wt.% $H_2O$, dynamic environmental and mixing conditions, but very different final physicochemical compositions due to different feedstock sequencing. Note that whilst the right activator

solution precipitates into a non-stable and highly viscous slurry, the left remains relatively clearer and stabilised with a better flowability, which is better combined with a geopolymer precursor for effective synthesis.

## 2 Experimental Procedure & Mathematical Model

### 2.1 Feedstock Materials and Calorimetry

For the calorimetry experiments, the temperature was measured using two k-type thermocouples encased in a stainless-steel sheath with a Tenmars TM-747D (0.1°C resolution and ± 0.05% relative uncertainty) 4-channel thermometer for data logging. A calorimetry setup was assembled using an IEC HL0820 cup calorimeter set consisting of a cupronickel rolled edge cylindrical insert within an insulated styrene foam cup encased in an outer cupronickel casing. A custom lid was created to fit the calorimeter cup assembly with a 'tower' so the temperature probe would sit immobile during magnetic stirring experimentation and maintain the inserted probe's constant height between different solutions. An ambient temperature thermocouple was positioned at a constant height beside the calorimeter. To maintain a consistent methodology of adding feedstocks to the calorimeter during in-situ experimentation, a port was added above the insert with a threaded cap and angled funnel (to allow for seamless and repeatable addition without making contact with the temperature probe), as seen in Figure 3 with all associated physical constants located in Appendix A [30].

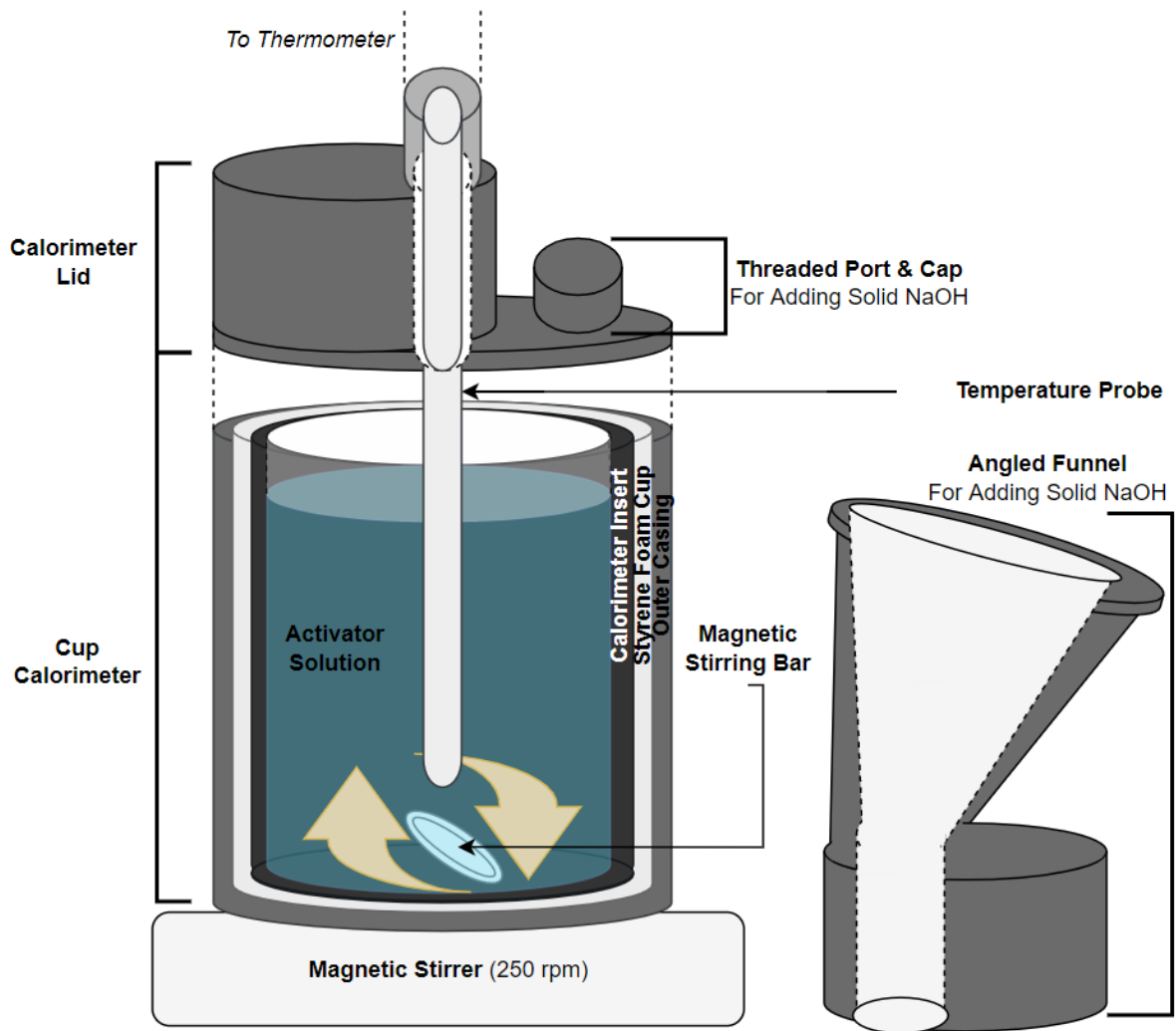

**Figure 3:** Experimental calorimeter setup used in this study.

The 115 mL activator solutions were prepared by first making up a 50 vol.% solution of predetermined caustic concentrations (e.g. 6M, 8M, 10M and 12M NaOH, etc.) followed by the remaining volume addition of sodium silicate ( aqueous sodium silicate with 30.4 wt.% $SiO_2$, 16.6 wt.% NaOH, 1.5 SG, from Coogee Chemicals Pty Ltd) to varying molar $SiO_2/Na_2O$ ratios representing a range of conventional activator solutions used in creating geopolymer composites. The caustic solutions were first prepared by weighing out (± 0.005% relative uncertainty) the required amount of solvent deionised water ($m_{H2O}$) within the insert containing a small PTFE magnetic stirring bar, placing the insert within the calorimeter assembly upon a magnetic stirrer set to 250 rpm, and beginning the data logging process with the thermometer. Once the mixing solvent had reached thermal equilibrium with the surroundings, the required pre-weighed amount of solid NaOH pellets (99% NaOH, Sigma-Aldrich) were added to the system via the angled funnel apparatus whilst simultaneously recording the timestamp of the three distinct feedstock addition sequences (i.e. threaded cap removal, addition of feedstock in the funnel to the system and replacing of the cap). Following the exothermic dissolution of the NaOH pellets ($m_{NaOH}$), the initial temperature of pre-weighed sodium silicate ($m_{SS}$) was measured with a separate thermocouple and routinely added to the system at 90% of the maximum temperature reached by the caustic solution. This convention was chosen to standardise experiments under the assumption of complete NaOH dissolution whilst minimising heat loss over time, as no consistent standard exists in the literature. Table 1 details the prepared solutions, accounting for NaOH purity and density in molarity calculations.

**Table 1:** Activator Solution Design Compositions

| Experiment/ Solution ID | NaOH Solution | Final Activator Solution | | |
|---|---|---|---|---|
| | $C_{NaOH}$ [M] | $m_{SS}$ [g] | $C_{NaOH}$ [M] | Molar $SiO_2/Na_2O$ |
| A | 14 | 85.6 | 10.1 | 0.7 |
| B | 12 | 85.6 | 9.1 | 0.8 |
| C | 10 | 85.6 | 8.1 | 0.9 |
| D | 8 | 85.6 | 7.1 | 1.0 |
| E | 6 | 85.6 | 6.1 | 1.2 |
| F | 4 | 85.6 | 5.1 | 1.4 |
| G | 3 | 85.6 | 4.6 | 1.6 |
| H | 2 | 85.6 | 4.1 | 1.8 |
| SS* | - | - | 6.3 | 2.4 |

* Pure Sodium Silicate ("SS") Feedstock for reference

## 2.2 Activator Solution Energy Balance & Dynamic Model

To quantify the energetics in a given activator solution system, a series of modelling equations, procedural statements and user-specified manipulated variables (i.e. mass of alkaline feedstocks, initial temperatures and solubility limits (etc.)) were entered and solved simultaneously with the Engineering Equation Solver software package (EES) [31]. EES was utilised to simulate numerous dynamic simulations to predict various thermodynamic properties of the geopolymer activator solution systems. All model variables are listed in Table A.1 (Appendix A) with standard units used.

The specific heat capacity for the activator solution ($C_{p,soln}$) was determined using an additive model incorporating the sequential addition of alkali and silicate feedstocks (Equation 1). Sequential addition of the sodium silicate feedstock was represented by the piecewise function $\omega(t)_{SS}$ in Equation 2, where $t_{SS,i}$ denotes the recorded addition time. The model proposed by Schrödle et al. [32] was used to describe the heat capacity for the alkali solution ($C_p^{NaOHSoln}$) with mass $m_{NaOHSoln}$, while the sodium silicate heat capacity, $C_p^{SS}$, was derived from the model proposed by Richet [33]. Solution density was approximated from literature [34], with conventional thermophysical constants (e.g. water heat capacity as a function of temperature etc.) defined using EES' in-built functions.

$$C_{p,soln} = \left[\left(C_p^{NaOHSoln} \cdot \frac{m_{NaOHSoln}}{m_{soln}}\right) + \omega(t)_{SS}\left(C_p^{SS} \cdot \left(1 - \frac{m_{NaOHSoln}}{m_{soln}}\right)\right)\right] \quad (1)$$

$$\omega(t)_{SS} = \begin{cases} 1, t_{SS,i} \geq t \\ 0, t_{SS,i} < t \end{cases} \quad (1)$$

The specific heat capacity of the full calorimeter system ($C_{p,System}$) was described by the summation equation of all heat capacities from the internal activator solution and immersed temperature probe to the outermost layer (with layers denoted as *j*) of the calorimeter multiplied by its specific weight as per Equation 3 below.

$$C_{p,System} = \sum_{j=1}^{n_j} \left(C_p^j \cdot \frac{m_j}{m_{System}}\right) \quad (2)$$

A heat balance equation can be applied to underpin the model whereby the change in energy to the system defines the solution's enthalpy change ($\frac{dH_{soln}}{dt}$) as defined in Equation 4. This equation can be rearranged to determine the solution temperature ($T_{soln}$) over a given time interval ($t_i$ and $t_f$, respectively). The system is referenced to the initial temperature ($T_{soln,i}$) of the activator solution and sodium silicate feedstock ($T_{SS,i}$), and with a known mass ($m_{soln}$) and heat capacity. The mathematical model utilised in this study was designed generally for ease of extrapolation to activator solution systems with differing mixing conditions, vessel geometries and temperature probes, which may vary depending on the place of batching and availability of the equipment to the user.

$$\frac{dH_{soln}}{dt} = \left(\sum \dot{Q}_{in} - \sum \dot{Q}_{out}\right)_{soln} = m_{soln}C_{p,soln}\frac{dT_{soln}}{dt}$$

$$\therefore T_{soln} = T_{soln,i} + \omega(t)_{SS}(T_{soln,i} - T_{SS,i}) + \int_{t_i}^{t_f}\left(\frac{\frac{dH_{soln}}{dt}}{(mC_p)_{soln}}\right)dt \quad (4)$$

### 2.2.1 Mass Transfer & Chemical Kinetics

Mass transfer and chemical kinetic models were used to quantify the driving changes of an analyte's concentration, particularly dissolved NaOH, to better estimate system energetics over time and compare with experimental data. The solid NaOH pellets were assumed to be completely spherical with a specific initial radius $r_{pellet,i}$ and volume, $V_{pellet,i}$ defined in Equations 5 and 6 respectively, where $N_{pellet,i}$ represents the number of pellets per unit mass. Conventional physicochemical relationships such as density, weight and molar fraction equations were applied throughout the model. The surface area of the solution, temperature probe and calorimeter layers were modelled as cylindrical layers, with solution mass and height ($h_{soln}$) varying as feedstocks were added.

$$V_{pellet,i} = \frac{4}{3}\pi r_{pellet,i}^3 = \frac{V_{NaOH,i}}{N_{pellet,i}} = \frac{(m_{NaOH}/\rho_{NaOH(s)})}{N_{pellet,i}} \tag{3}$$

$$r_{pellet} = r_{pellet,i} + \omega(t)_{NaOH} \int_{t_i}^{t_f} \left(\frac{dr_{pellet}}{dt}\right) dt \tag{6}$$

As the solute pellets are dissolved by water in solution at the boundary layer defined by $r_{pellet}$ with a surface area of $SA_{pellet}$, their specific mass is transitioned from an initial solid state at saturation (with a solubility concentration $C_{NaOH}^{sat}$) to the mass associated with the general concentration of the aqueous solution ($C_{NaOH(aq)}$). This process, eventually leading to the complete dissolution of the NaOH pellets (i.e. when $r_{pellet} = 0$) is characterised by the differential change in radius and volume of the pellets over time as conveyed by the modified Nernst-Brunner equation [35], with a mass transfer dissolution constant $k_{NaOH,diss}$ and boundary layer contextual parameter $\omega(t)_{NaOH}$, as per the Equations 7 and 8 below. These dynamics allow for estimating the pellets dissolved in Equations 9 and 10.

$$\omega(t)_{NaOH} = \begin{cases} 1, r_{pellet} > 0 \\ 0, r_{pellet} \leq 0 \end{cases} \tag{7}$$

$$\frac{dr_{pellet}}{dt} = -k_{NaOH,diss}\left(C_{NaOH}^{sat} - C_{NaOH(aq)}\right) \tag{8}$$

$$\frac{dV_{pellet}}{dt} = SA_{pellet}\pi \frac{dr_{pellet}}{dt} \tag{9}$$

$$V_{pellet} = V_{pellet,i} + \omega(t)_{NaOH} \int_{t_i}^{t_f} \left(\frac{dV_{pellet}}{dt}\right) dt \qquad (10)$$

Mass transfer from the liquid sodium silicate ($\dot{m}_{SS}$) was assumed to occur more simplistically as the instantaneous mixing of the liquid inversely proportional to the addition interval (i.e. when the flowing liquid is gradually added to the alkali solution at time $t_{SS,i}$ to $t_{SS,f}$). Equation 11 describes this dynamic where $k_{SS}$ is the dimensionless mass transfer constant of sodium silicate.

$$\dot{m}_{SS} = \omega(t)_{SS} \left(k_{SS} \frac{m_{SS}}{t_{SS,f} - t_{SS,i}}\right) \qquad (11)$$

Solving these equations simultaneously yields expressions for the dissolved NaOH concentration within the activator solution at any given time with Equations 12-13 (where $M_{NaOH}$ is the molar mass of NaOH, $\dot{n}_{NaOH(aq)}$ is the NaOH molar flow and $m_{SS,NaOH}$ is the NaOH mass within the sodium silicate).

$$\dot{n}_{NaOH(aq)} = \frac{\left(-\frac{dV_{pellet}}{dt}\rho_{NaOH(s)}N_{pellet,i}\right) + \left(\dot{m}_{SS}\frac{m_{SS,NaOH}}{m_{SS}}\right)}{M_{NaOH}} \qquad (4)$$

$$C_{NaOH(aq)} = \frac{\int_{t_i}^{t_f} \dot{n}_{NaOH(aq)} dt}{V_{soln}} \qquad (5)$$

### 2.2.2 Energy Outputs

Heat loss from the activator solution was assumed to occur via conduction ($\dot{Q}_{Cond,Sys}$) driven by the temperature gradient between the solution and each subsequent thermal layer, and via convection ($\dot{Q}_{Conv,Sys}$) from the outermost layer ($T_{Surr}$) to the surrounding ambient temperature ($T_{Amb}$). Heat transfer between the activator solution and the temperature probe ($\dot{Q}_{Probe}$, with temperature $T_{Probe}$) is described in Equations 14-17, with additional solution transfers outlined in Appendix B. The probe's time constant ($\tau_{Probe}$) and system's heat transfer coefficient ($U_{Sys}$) were determined empirically from experimental equipment and data (details in Appendix B).

$$\dot{Q}_{Cond,Sys} = U_{Sys} SA_{Sys} (T_{Soln} - T_{Surr}) \qquad (14)$$

$$\dot{Q}_{Probe} = U_{Probe} SA_{Probe} (T_{Soln} - T_{Probe}) \qquad (15)$$

$$\tau_{Probe} = \frac{m_{probe} C_{p,Probe}}{U_{Probe} SA_{Probe}} \tag{16}$$

$$\therefore T_{Probe} = T_{soln,i} + \int_{t_i}^{t_f} \dot{Q}_{Probe}\, dt \tag{17}$$

Heat transfer arising from the addition of sodium silicate to the system ($\dot{Q}_{SS}$) was modelled as a disturbance to the solution's temperature as per Equation 18. This involves heat loss from the solution to the added sodium silicate mass between the initial and final addition periods at its initial temperature. Therefore, from the equations listed above (and supplementary in Appendix B), the total heat loss from the solution can be modelled as per Equation 19.

$$\dot{Q}_{SS} = \omega(t)_{SS} \left( \dot{m}_{SS} C_{p,SS} (T_{Soln} - T_{SS,i}) \right) \tag{18}$$

$$\dot{Q}_{Out,soln} = \dot{Q}_{Cond,Sys} + \dot{Q}_{Probe} + \dot{Q}_{SS} + \dot{Q}_{Conv,Mixer} \tag{19}$$

### 2.2.3 Energy Inputs

In activator solutions where, no external heating is applied, heat input is assumed to come solely from feedstock additions at their initial temperatures and user-specified times, along with any associated thermodynamic processes associated with each feedstock. For solid NaOH pellets, a lumped enthalpy term ($\Delta H_{NaOH(s)}$) was used and assumed to account for all thermophysical effects, including exothermic dissolution and endothermic hydrate formation (NaOH·nH$_2$O). The dominant enthalpy change in the NaOH solution arises from the exothermic heat of dissolution, which is known to change as a function of concentration due to hydration energy differences [36]. In dilute solutions ($\leq$ 6 M), excess water fully hydrates the Na$^+$ and OH$^-$ ions, and the heat of dissolution is assumed constant from literature at -44.5 kJ/mol [37]. However, in concentrated solutions (> 6 M), fewer water molecules relative to the number of ions results in incomplete hydration and a lowering of the effective heat of dissolution with increasing concentration.

Calorimetric experiments were conducted to quantify NaOH solution enthalpy changes at varying concentrations model validation. For each concentration, the solution's enthalpy change ($\Delta H_{soln}$), was averaged from three experiments and compared with literature values in the results section [28, 38]. In concentrated solutions, these values were combined with the calorimeter's enthalpy change to determine the total system enthalpy change ($\Delta H_{NaOH(s),exp}(C_{NaOH(aq)})$), used as the heat of reaction at higher concentrations. The enthalpy of reaction is defined by the piecewise Equation 20, distinct for dilute and concentrated solutions, with further details in Appendix B.

$$\Delta H_{NaOH(s)} = \omega(t)_{NaOH} \begin{cases} -44.5, & C_{NaOH(aq)} \leq 6\,M \\ -\Delta H_{NaOH(s),exp}(C_{NaOH(aq)}), & 6\,M < C_{NaOH(aq)} \leq 14\,M \end{cases} \quad (20)$$

Enthalpy changes from sodium silicate addition ($\Delta H_{SS(aq)}$) were attributed to dilution enthalpy and equilibration between differing NaOH concentrations in the added feedstock and NaOH solution as defined in Equation 21. The enthalpy of dilution ($\Delta H_{NaOH,Dilution}$) was incorporated by using the model from Simonson et al. [39]. Based on this, the total heat input ($\dot{Q}_{in,soln}$) from the activator solution is modelled by Equation 22.

$$\Delta H_{SS(aq)} = \omega(t)_{SS} \cdot \Delta H_{NaOH,Dilution} \quad (21)$$

$$\dot{Q}_{in,soln} = \dot{n}_{NaOH(aq)}(\Delta H_{NaOH(s)} + \Delta H_{SS(aq)}) \quad (22)$$

## 2.3 Activator Solution Thermodynamic Stability

The thermodynamic stability of activator solutions is defined as the time for the solution to reach a steady state enthalpy after the addition of a given feedstock (within a specified tolerance, $\varepsilon = 1\,J/s$). Stability points, $t_{stable,NaOH}$ and $t_{stable,SS}$, corresponding to after the NaOH solution reaches its maximum temperature ($t_{T_{Max}}$) and sodium silicate addition, are given by Equations 23-24. Theoretically, $t_{stable,NaOH}$ should coincide with the sodium silicate addition time. However, since this point was unknown before experimentation, it was assumed to be 90% of the maximum temperature ($T_{Max}$) reached by the caustic solution, as mentioned previously. The overall thermodynamic stability point, $t_{stable}$, calculated in Equation 26 represents the earliest time at which the solution is stable, aiding in the optimisation of batching times for geopolymer systems.

$$t_{stable,NaOH}(t) = \min\left\{t: \left|\frac{dH_{soln}}{dt}\right| \leq \varepsilon\ \forall\ t \geq t_{T_{Max}}\right\} \quad (23)$$

$$t_{stable,SS}(t) = \min\left\{t: \left|\frac{dH_{soln}}{dt}\right| \leq \varepsilon\ \forall\ t > t_{SS,i}\right\} \quad (24)$$

$$t_{stable} = t_{stable,NaOH} + t_{stable,SS} \quad (25)$$

## 3  Results & Discussion

### 3.1  Thermochemistry & Temperature

Figure 4 presents the first set of NaOH dissolution experiments, illustrating the enthalpy change of solution results for comparison with literature. These results allow for the more accurate derivation of the NaOH reaction enthalpy as a function of concentrated solutions between 6-14 M NaOH, which is conveyed in Table 2. Experimental values (which were the measured feedstock masses for better accuracy) align well with literature data on stirred NaOH solutions [28, 38]. While minor deviations from Bespalko et al. (overlayed in Figure 4) emerge at moderate concentrations (≥ 6 M NaOH) and increase slightly at higher, nonideal concentrations, they remain within the error bars – calculated as the standard error with a 95% confidence interval from the mean of three experiments – and are considered valid for this model. These deviations likely stem from differences in calorimeter materials, stirring speeds, and NaOH feedstock purity, particularly unmeasured hygroscopicity. Comparisons with other sources show agreement in bulk molar enthalpy [kJ/mol], though direct comparison is challenging due to unspecified details on equipment materials (i.e. the calorimeter used) and thermodynamic parameters required to accurately determine specific enthalpies of the solution components [38].

In this study, the NaOH enthalpy of solution variable combined dissolution and hydration effects. Various NaOH hydrates (i.e. NaOH·$n$H2O where $n$ has been found in various studies to equal 1, 2, 3, 3.11, 3.5, 4, 5, 7) exist in solution, each with distinct heat capacities and formation enthalpies that influence dissolution enthalpy [40, 36, 41, 38]. However, the distribution of the $n ≥ 2$ hydrates predominantly form at temperatures lower than 25 °C (and more abundantly at T ≤ 0°C), and are assumed to have negligible impacts in this study. The NaOH monohydrate ($n = 1$) remains stable up to 65 °C at concentrations less than ≈ 10 wt.% NaOH but gradually dissociates with higher temperatures as equilibrium shifts to form anhydrous NaOH [42]. While its specific enthalpy of formation was not explicitly parameterised in the model equations, the lumped heat of reaction described in Equation 26 ($R^2 = 0.99$, used in Equation 20) indirectly accounts for its endothermicity. This approach captures the nonideal thermochemical behaviour of NaOH dissolution, preventing unrealistic model predictions (e.g., maximum temperatures exceeding the boiling point of concentrated solutions) that would result from assuming a constant heat of dissolution (−44.5 kJ/mol).

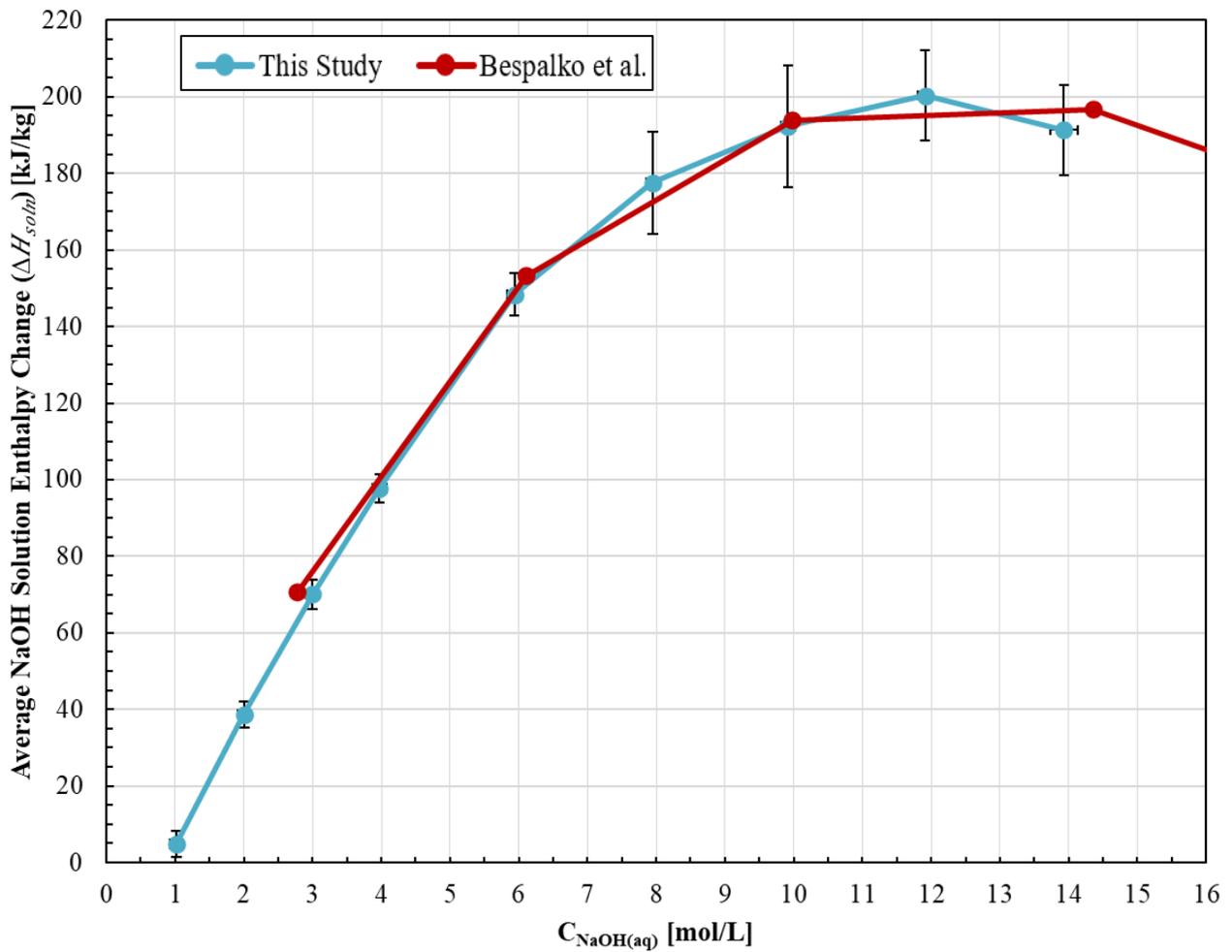

**Figure 4**: Experimental NaOH enthalpy of solution change (before addition of sodium silicate) compared with literature results of Bespalko et al. [28]. Values for this study represent the mean ± 95% confidence interval.

**Table 2:** NaOH concentration and heat of reaction results. Experimental values are presented as the mean ± 95% confidence interval, rounded to 1 dp.

| Experiment/ Solution ID | NaOH Solution | | |
|---|---|---|---|
| | Design $C_{NaOH(aq)}$ [M] | Experimental $C_{NaOH(aq)}$ [M] | $\Delta H_{NaOH(s)}$ [kJ/ mol] |
| A | 14 | 13.9 ± 0.2 | -29.0 ± 1.4 |
| B | 12 | 11.9 ± 0.1 | -33.6 ± 1.4 |
| C | 10 | 9.9 ± 0.1 | -36.9 ± 2.2 |
| D | 8 | 7.9 ± 0.1 | -40.4 ± 2.2 |
| E | 6 | 5.9 ± 0.1 | -42.8 ± 1.3 |
| F-H | - | < 6 | -44.5* |

*Assumed NaOH enthalpy constant from literature for dilute solutions [37].

$$\Delta H_{NaOH(s),exp}\big(C_{NaOH(aq)}\big) = 0.0749 C_{NaOH(aq)}^2 + 0.2223 C_{NaOH(aq)} - 46.691 \qquad (26)$$

Figure 5 presents the dynamic temperature profiles for B, D, and E activator solutions (note that other profiles were omitted from the figure for ease of viewing) with data normalised to the NaOH addition time. All profiles exhibit a sharp temperature increase after the addition of NaOH(s) due to its exothermic dissolution into solution (forming NaOH(aq)), followed by a period of decaying temperature (i.e. heat loss from the system) and a distinct decrease of 14-16 °C upon addition of the cooler sodium silicate. The experimental (circle data points) and modelled (solid black lines) temperature responses generally align well along their profiles, with minor deviations between them observable. The steepness and slight misalignment of the initial temperature rise seem proportionate to the concentration arising from additional solute dissolution, with larger gaps between measured data points occurring from the resolution of the temperature probe. As solutions of higher concentrations reach their respective temperature maximums sooner, the model appears to predict these maximums with small deviations accurately.

Temperature deviations of 1.5 - 2 °C in progressively higher concentrated solutions at 5 minutes, such as in Solution B, occur due to their more nonideal behaviour, contrasted with smaller deviations in lower concentrations (e.g. 1.1 °C in the as part of Solution E). The proportionality between maximum temperature deviations and concentration is attributed to the increasingly complex solute-solute intermolecular interactions present with transient dynamics that are not encompassed from the dissolution equations in this model. Additionally, a 5-10 °C mismatch occurs between the experimental and modelled trends as solutions quickly approach their maximum temperatures between 0.5 - 0.75 minutes and is proportional to concentration. This discrepancy likely stems from the model assuming uniformly spherical and evenly dissolvable NaOH pellets. In practice, the pellets are slightly elliptical, of slightly different unit masses and likely have some minor heterogeneity in hygroscopic water unaccounted for in the model. The model follows a largely linear decay after reaching a solution's maximum temperature but slightly diverges for higher-concentration solutions (e.g., B).

While boiling points increase with NaOH concentration (e.g., ≈ 120°C for Solution A), a better fit would likely be achieved by including the energy lost embodied in the mass of particles that leave as vapour [40]. At 90% of the maximum temperature, sodium silicate addition causes the modelled equations to ramp down to a new steady state temperature, with ± 2 °C deviations at higher concentrations due to additional alkali-silicate interactions. Notably, both experimental E and D trends show a sharp temperature pulse after the addition around 8.6 minutes, whereas the B experiment does not. Whilst sodium silicate's dilution enthalpy has been accounted for in the model, this pulse likely arises from exothermic effects as sodium equilibrates with the silica species, breaking bonds and altering solution speciation; associated in literature as sodium silicate's enthalpy of mixing [43]. In experiment B, little to no exothermic activity is observed to occur. However, the model is slightly less predictive in its resulting steady-state temperature. Additionally, this temperature pulse could partially result from viscosity-related heat dissipation delays, as sodium silicate viscosity decreases at higher temperatures [44]. While sodium silicate's dissolution exothermicity is beyond this study's scope, future research is recommended.

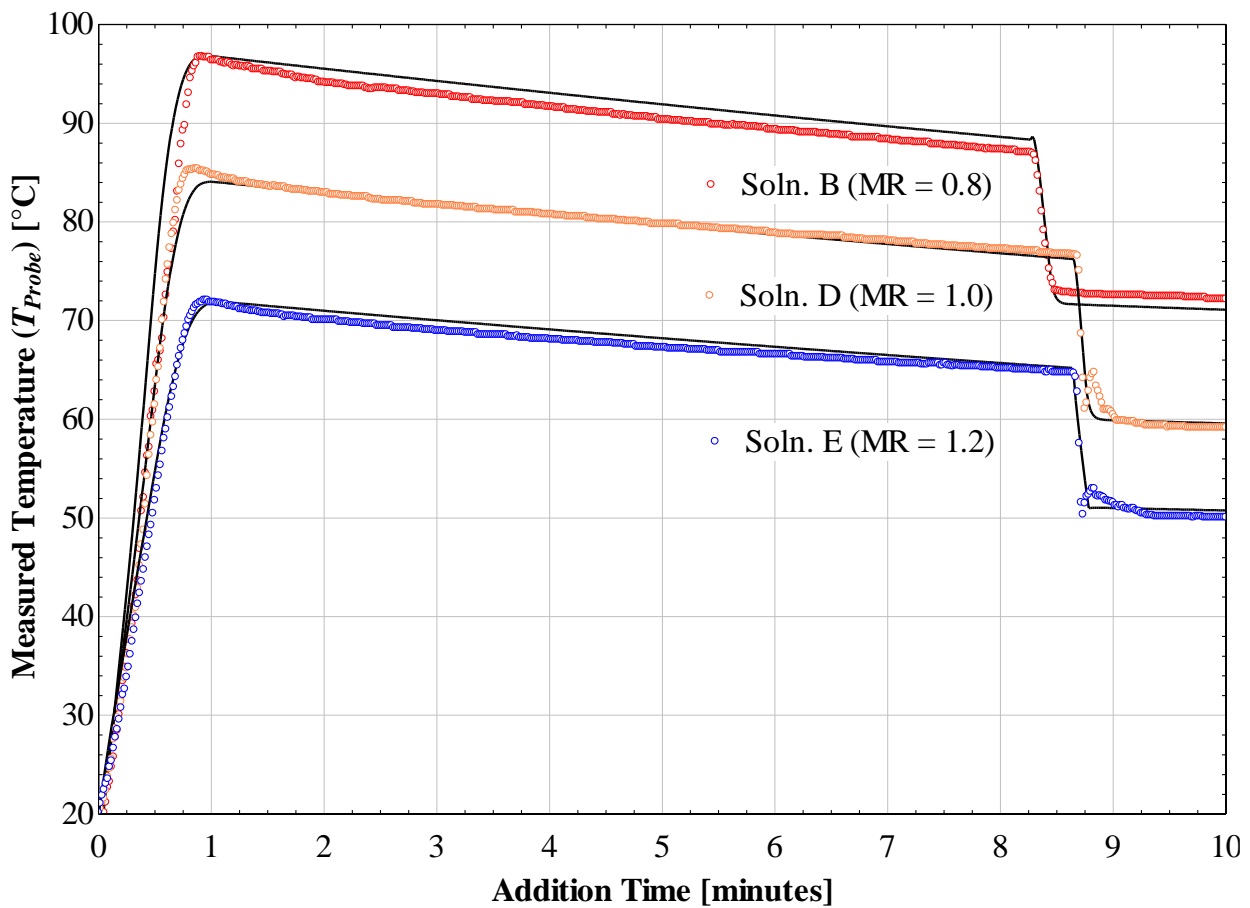

**Figure 5**: Temperature profiles for the B, D and E activator solution experiments with coloured circles and solid black lines representing the experimental measurements and relative model outputs respectively.

## 3.2 Thermodynamic Stability

Figure 7 illustrates how the solution's enthalpy change can be used to identify key temperature and stability milestones during activator solution preparation. In Experiment E, the differential profile (solid purple line) indicates thermodynamic stability at values approaching 0 J/s, where enthalpy changes arise solely from heat loss rather than bulk chemical activity. Deviations from this point signifies instability due to ongoing chemical activity, primarily from the NaOH exothermic dissolution and sodium silicate addition at 90% of the solution's maximum temperature at ≈ 8.7 minutes. Minimal deviation from the zero-line suggests that after this point, heat changes result only from system heat loss, reinforcing the activator solution's thermostability. Although sodium silicate was consistently added at $90\%T_{Max}$ for experimental consistency, the stability profile in Figure 7 suggests that the solution's enthalpic stability (at least at a bulk level) is reached much earlier. Quantitative results indicate that sodium silicate could have been added as early as 1.29 minutes, rather than at 8.7 minutes (experimental) or the 24-hour delays sometimes cited in literature (discussed previously). Table 2 presents stability metrics for selected activator solutions. All trials exhibit a short instability period after sodium silicate addition between 7 - 12 seconds, compared to NaOH dissolution, which ranges from 0.9 - 1.2 minutes. For Experiment E, the results reveal that the solution can be considered stable and ready to use within 1.3 minutes, and not the original experimental assumption of 8.9 minutes nor from what is conventionally done in literature. The calculated stable-batching times for activator solutions demonstrate that activator solutions can be ready and stable in a matter of minutes – a stark contrast to literature reported times of up to (or exceeding) 24 hours.

After sodium silicate addition, the differential enthalpy responses between the model and experiment diverge along the addition interval (i.e. between $t_{SS,i}$ and $t_{SS,f}$). Deviations occur due to the model's poor approximation of the sodium silicate's exothermicity impacting the temperature, as discussed above. Despite this, Experiment E rapidly reaches a steady state temperature with no observable differential enthalpy change, supporting the notion of bulk thermochemical stability. This provides a quantifiable event for batching the activator solution in a geopolymer system. When considering scale-up applications of geopolymer manufacturing, adopting this standardised method could save hours of operational time and cost, particularly in designing geopolymer activator reactors with shorter residence times and batching process tolerances.

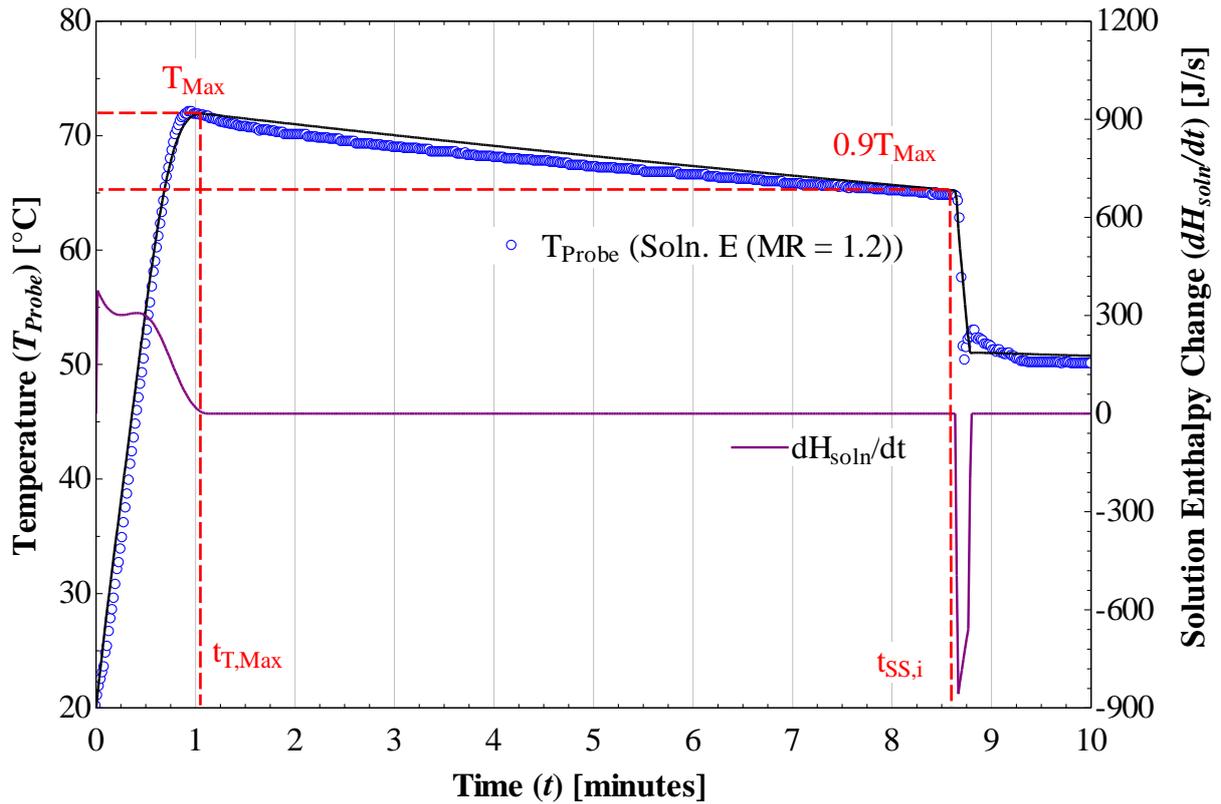

Figure 7: Temperature and differential enthalpy profiles for the experimental E activator solution with blue circles and solid black and purple lines representing the experimental measurements and model outputs for the temperature and solution enthalpy, respectively. Red dashed lines are added to clarify the different thermodynamic milestones within the activator solution.

Table 3: Comparison of the B, D and E experimental, model and literature Activator solution thermostability Metrics (to 2 decimal places).

| Experiment/ Solution ID | Experimental Activator Solution | | Activator Solution Stability Times [mins] | | | Literature[*] |
|---|---|---|---|---|---|---|
| | Molar $SiO_2/Na_2O$ | $t_{Stable}$ [mins] (This Study, Experimental) | $t_{Stable,NaOH}$ = $t_{T_{Max}}$ | $t_{Stable,SS}$ | $t_{Stable}$ (This Study, Calculated) | $t_{Stable}$ [hours] |
| B | 0.8 | 8.67 | 0.92 | 0.19 | 1.11 | ≥ 24 |
| D | 1.0 | 8.80 | 0.99 | 0.12 | 1.11 | ≥ 24 |
| E | 1.2 | 8.85 | 1.14 | 0.15 | 1.29 | ≥ 24 |

*Literature "stability" times (or points where the activator solution is used in batching) which, when specified, can range from 1 – 24 hours. However, most times are unspecified or inconsistent (e.g. "mixed overnight").

## 3.3 Activator Solution Batching Optimisation and Recommendations for Future Work

Depending on the situation, the batching time and sequencing of the stabilised activator solution may be time-sensitive such that it can be used sooner (i.e. for industrial purposes where operational expenditures (OPEX) are desirably minimised), later (i.e. allowing to cool to ensure no mass loss when forming and curing geopolymer composites) or at certain points for research and development in the laboratory or quality control monitoring. Figure 8 provides a guideline for estimating when an activator solution with a given $SiO_2/Na_2O$ ratio cools to a desired temperature for use. These profiles can be adjusted by modifying heat transfer conditions, such as vessel properties and solute dissolution methods (solid vs. liquid feedstocks). This framework supports process scale-up, defining operational limits for heat transfer, maximum temperatures, and safety considerations (e.g., handling hot corrosive solutions). The time to cool to ambient temperature decreases with higher molar $SiO_2/Na_2O$ because the sodium silicate is added to a solution with lower NaOH concentration, which retains less heat. This results in a more significant temperature drop and faster cooling. In contrast, solutions with lower $SiO_2/Na_2O$ (higher NaOH concentration) have greater thermal mass, causing them to cool more slowly, despite their larger initial temperature difference.

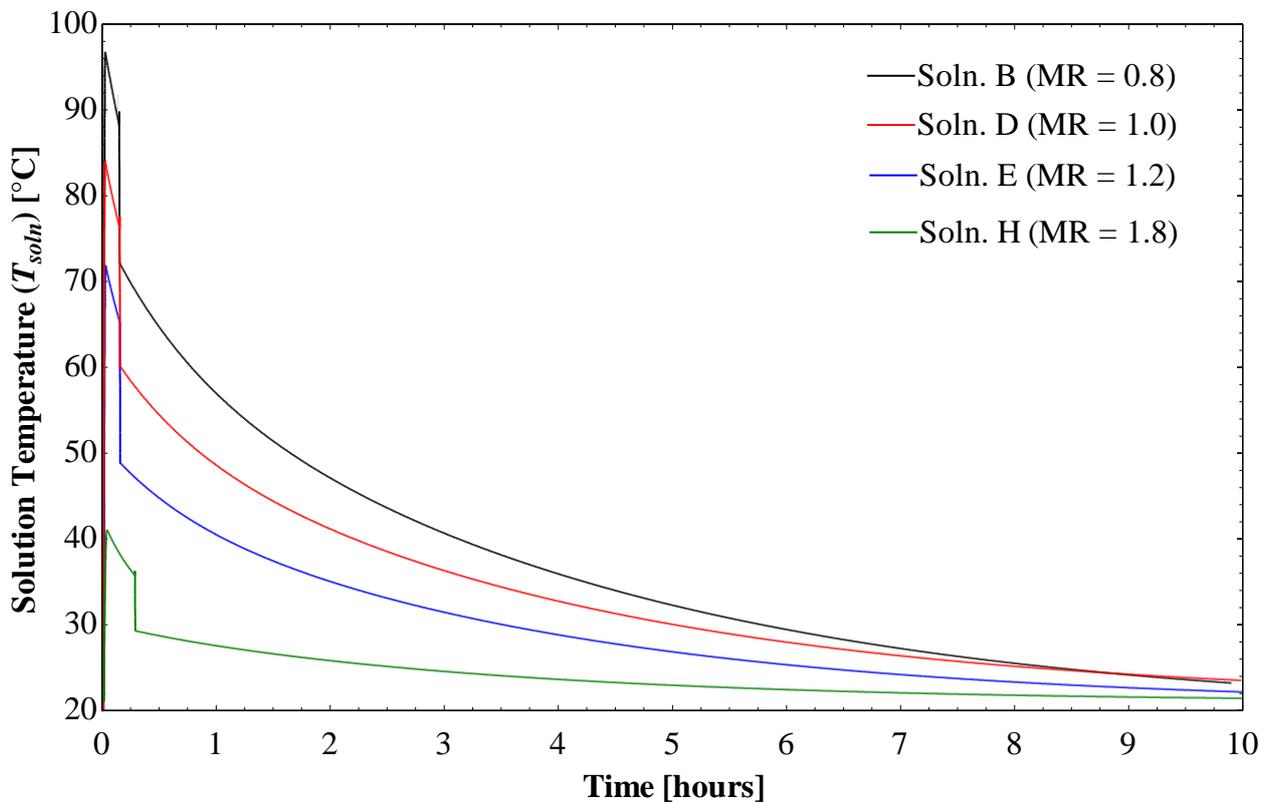

**Figure 8**: Modelled cooling time of activator solutions ("soln.") with labelled characteristic molar $SiO_2/Na_2O$ ratios ("MR").

While the model aligns well with experimental data, future work could enhance accuracy by incorporating quantitative nuclear magnetic resonance spectroscopy (qNMR) to track silica speciation, solubility, and its precipitation dynamics. This would further enhance the model's applicability and allow for a greater understanding of activator solution physicochemical stability. Further research on sodium silicate's exothermicity at different mixing ratios would refine predictions. Comparative experiments on geopolymer systems with different precursors (e.g., fly ash and metakaolin) to assess mechanical and durability properties, alongside investigations into correlations between qNMR-derived activator speciation, X-ray diffraction (XRD) amorphous characterisation of solid precursors, and final geopolymer properties, would further optimise activator formulations and enhance overall geopolymer performance. Expanding these studies to include improved heat transfer models based on vessel geometry, alternative alkali cations (i.e. potassium-based activator solutions) and commercially available (aqueous) sodium hydroxide solutions could help standardise geopolymer synthesis across academic and industrial settings and broaden applications of geopolymer technology.

## 4    Conclusions

This study presents a mathematical model for the dynamic thermophysical assessment of geopolymer activator solutions, offering a quantitative framework to predict temperature profiles, NaOH reaction enthalpies, cooling times, and thermodynamic stability. Validated against experimental data, the model is highly adaptable – allowing modifications in heat transfer coefficients, solution compositions, feedstocks, and vessel geometries – to accommodate both laboratory and industrial applications. The findings challenge conventional geopolymer synthesis methods, which often rely on extended and poorly defined equilibration periods of up to 24 hours. Instead, the model establishes that activator solutions can reach thermodynamic stability within a minute, significantly reducing batching times. This predictive capability not only enhances process efficiency but also aids in developing safety protocols by estimating maximum temperatures for handling hot, potentially corrosive activator solutions.

By integrating these insights into Standard Operating Procedures (SOPs), practitioners can optimise batching protocols for informed activator solution design to assess technical feasibility, improve quality control, optimise solution synthesis and advance the broader application of geopolymer technologies.

## 5 Acknowledgements

The authors acknowledge the facilities at Curtin University, specifically the John de Laeter Centre, for providing the laboratory for experimentation and storage of feedstocks and the staff at Coogee Chemicals Australia for assistance in procuring sodium silicate. Furthermore, the authors would like to extend an acknowledgement to Ching Yong Goh from the School of Molecular and Life Sciences at Curtin University for the provision of calorimetry equipment and finally to Hendrik Gildenhuys and Jarrad Allery for assistance in running the calorimetry experiments. This project has been undertaken as part of the Process Legacy subproject with financial support from the Future Battery Industries Cooperative Research Centre, established under the Australian Government's Cooperative Research Centres Program.

## 6 Conflicts of Interest

The authors declare that they have no known competing financial interests or personal relationships that could have appeared to influence the work reported in this paper.

# 8 Supplementary Information & Appendices

## 8.1 Appendix A: List of Variables and Notation

Refer below to the nomenclature used throughout this document where the following subscripts are used but omitted from the list below for conciseness:

- *i* and *f* are used with some symbols to indicate the initial and final states of those variables (e.g. $C_{NaOH,i}$ and $C_{NaOH,f}$).
- *sys* and *j* are used to characterise the full system and its sub-components respectively.
- *x* to denote different feedstocks (i.e. NaOH pellets, solvent water or sodium silicate).

| Symbol | Parameter Definition | [Units] |
|---|---|---|
| $C_{NaOH(aq)}$ | Final Concentration of NaOH | mol/L |
| $C_{NaOH}^{sat}$ | Unstable (Saturation) Concentration of NaOH | mol/L |
| $C_p^x$ | Heat capacity of component *x* | $\frac{J}{g \cdot °C}$ |

| Symbol | Description | Units |
|---|---|---|
| $C_p^j$ | Heat capacity of calorimeter layer $j$ | $\frac{J}{g \cdot °C}$ |
| $\Delta H_{Exp}$ | Enthalpy change of experimental sample set | kJ/mol |
| $\Delta H_{NaOH(s),exp}(C_{NaOH(aq)})$ | Experimentally derived NaOH Heat of dissolution | kJ/mol |
| $\Delta H_{NaOH(s)}$ | NaOH Heat of dissolution | kJ/mol |
| $\Delta H_{NaOH,Dilution}$ | NaOH Heat of dilution | kJ/mol |
| $\Delta H_{soln}$ | Enthalpy of Activator Solution | kJ/kg |
| $\Delta H_{SS(aq)}$ | Enthalpy Change from Sodium Silicate addition | kJ/mol |
| $\frac{dH_{soln}}{dt}$ | Enthalpy of Activator Solution Change | J/s |
| $\frac{dr_{pellet}}{dt}$ | Change in radius of NaOH pellet | mm/s |
| $\frac{dT_{soln}}{dt}$ | Differential Temperature of activator solution | °C/s |
| $\frac{dV_{pellet}}{dt}$ | Change in volume of NaOH pellet | mm³/s |
| $\varepsilon$ | Differential Enthalpy Tolerance | J/s |
| $h$ | Convective heat transfer coefficient | $\frac{W}{m^2 \cdot °C}$ |
| $h_{soln}$ | Height of activator solution | mm |
| $k_{NaOH,diss}$ | NaOH pellet mass transfer coefficient | m⁴/mols·s |
| $k_{SS}$ | Sodium silicate mass transfer coefficient | - |
| $m_j$ | Mass of component j | g |
| $m_x$ | Mass of feedstock x | g |
| $m_{SS,NaOH}$ | Mass of NaOH in sodium silicate feedstock | g |
| $\dot{m}_{SS}$ | Mass flow of sodium silicate | g/s |
| $M_{NaOH}$ | Molar Mass of NaOH | g/mol |
| $N_{pellet,i}$ | Unit NaOH pellet mass | pellet |
| $\dot{n}_{NaOH(aq)}$ | Molar flow of NaOH | mol/s |
| $\rho_x$ | Density of feedstock x | g/L |
| $Q_j$ | Heat transfer of component j | kJ/s |
| $Q_x$ | Heat transfer of feedstock x | kJ/s |
| $\dot{Q}_{Cond,j}$ | Conductive heat transfer of component j | kJ/s |
| $\dot{Q}_{Conv,j}$ | Convective heat transfer of component j | kJ/s |

| Symbol | Description | Units |
|---|---|---|
| $\dot{Q}_{SS}$ | Heat flow from sodium silicate addition | $kJ/s$ |
| $r_{pellet,i}$ | Initial radius of NaOH pellet | $mm$ |
| $r_{pellet}$ | Radius of NaOH pellet | $mm$ |
| $SA_j$ | Surface Area of component $j$ | $mm^2$ |
| $T_{amb}$ | Ambient Temperature | $°C$ |
| $T_{Max}$ | Activator solution temperature maximum | $°C$ |
| $T_{Probe}$ | Temperature of Probe | $°C$ |
| $T_{SS,i}$ | Temperature of ambient sodium silicate | $°C$ |
| $T_{soln}$ | Temperature of activator solution | $°C$ |
| $T_{surr}$ | Temperature of surroundings / calorimeter edge | $°C$ |
| $T_{soln,i}$ | Initial temperature of the activator solution | $°C$ |
| $t_{T_{Max}}$ | Time at solution maximum temperature | $s$ |
| $t_{SS,f}$ | Final addition time of sodium silicate | $s$ |
| $t_{SS,i}$ | Initial addition time of sodium silicate | $s$ |
| $t_f$ | Final time interval | $s$ |
| $t_i$ | Initial time interval | $s$ |
| $t_{stable,NaOH}$ | Stability Time of Activator solution after NaOH addition | $s$ |
| $t_{stable,SS}$ | Stability Time of Activator solution after sodium silicate addition | $s$ |
| $t_{stable}$ | Stability Time of Activator solution after all feedstocks added | $s$ |
| $\tau_{Probe}$ | Measuring probe time constant | $s$ |
| $U_{Probe}$ | Overall probe heat transfer coefficient | $\frac{J}{s \cdot m^2 \cdot C}$ |
| $U_{Sys}$ | Overall system heat transfer coefficient | $\frac{J}{s \cdot m^2 \cdot C}$ |
| $V_{NaOH,i}$ | NaOH volume | $mm^3$ |
| $V_{pellet,i}$ | NaOH pellet initial volume | $mm^3$ |
| $V_{pellet}$ | NaOH pellet volume | $mm^3$ |
| $V_{soln}$ | Volume of solution | $mm^3$ |
| $\omega(t)_{NaOH}$ | Context piecewise function for NaOH feedstock addition | - |
| $\omega(t)_{SS}$ | Context piecewise function for sodium silicate feedstock addition | - |

Table A.1: Heat transfer constants used in this work, to 2 dp [30].

| Parameter | Heat Capacity ($C_p^j$) [J/(g·°C)] | Convective Heat Transfer Constant ($h$) [W/(m²·°C)] |
|---|---|---|
| Magnetic Stirring Bar (PTFE "Teflon") | 1.30 | - |
| Insert (Cupronickel) | 0.38 | - |
| Cup (Polystyrene) | 1.06 | - |
| Casing (Cupronickel) | 0.38 | - |
| Probe (Stainless Steel) | 0.46 | - |
| NaOH Solution Convection Constant (Liquid Forced Flow) | - | 2065 |
| Activator Solution Convection Constant (Liquid Forced Flow) | - | 350 |

## 8.2 Appendix B: Extended Modelling Equations

Within the solution, heat transfer between the activator solution, mixing apparatus – which in the case of this research paper referred to the magnetic stirring bar within the calorimeter – and the temperature measuring probe ($T_{Probe}$) is conveyed in the equations below where the probe's time constant ($\tau_{Probe} = 2\ s$) and system's heat transfer coefficient ($U_{Sys}$) is a user-specified variable and was derived empirically from experimental data. A series of overall heat transfer equations were used to describe heat dynamics between the solution and calorimeter.

$$\dot{Q}_{Conv,Mixer} = h_{Mixer} SA_{Mixer} (T_{Soln,f} - T_{Mixer}) \tag{B.1}$$

$$\dot{Q}_{Cond,Sys} = U_{Sys} SA_{Sys} (T_{Soln} - T_{Surr}) \tag{B.2}$$

$$\dot{Q}_{Conv,Sys} = h_{Sys} SA_{Sys} (T_{Surr} - T_{Amb}) \tag{B.3}$$

Additionally, the surrounding temperature at the outermost layer of the calorimeter with respect to the solution can be defined with the equation below.

$$T_{surr} = T_{amb} + \int_{t_i}^{t_f} \left( \frac{\dot{Q}_{out,soln} - \dot{Q}_{Conv,Sys}}{(mC_p)_{system}} \right) dt \tag{B.4}$$

The enthalpy change of reaction was calculated from the enthalpy change of solution and calorimeter experimental sample set ($\Delta H_{Exp}$) from the equations below.

$$\Delta H_{soln} = \Delta H_{soln,T_{Max}} - \Delta H_{soln,T_i} \tag{B.5}$$

$$\Delta H_{Exp} = \sum_{j=1}^{n_j} C_{p,T_i}^{j} (T_{Max} - T_i) \tag{B.6}$$

$$\Delta H_{NaOH(s)} = M_{NaOH} \Delta H_{soln} + \Delta H_{Exp} \tag{B.7}$$